\documentclass[11pt]{article}

\usepackage[T1]{fontenc}
\usepackage{floatrow}
\usepackage[greek,english]{babel}
\usepackage{amsthm}
\usepackage{authblk}
\usepackage[sort&compress]{natbib}
\usepackage{epigraph}
\usepackage{graphicx}
\graphicspath{{.}}
\usepackage{comment}
\usepackage{textgreek}
\usepackage{bm}
\usepackage[T1]{fontenc}
\usepackage{xcolor}
\definecolor{oxblue}{rgb}{0,0.1294117647,0.27843137254} 
\definecolor{forestgreen}{HTML}{008000}

\usepackage[hidelinks,colorlinks=true]{hyperref}
 \hypersetup{
 colorlinks=true,
 linkcolor=blue,
 filecolor=blue,
 citecolor=blue, 
 urlcolor=blue
 }
\usepackage{fullpage}
\usepackage[right]{lineno}
\usepackage{amssymb}
\usepackage{mathtools}
\usepackage{amsmath} 
\usepackage{graphicx}
\graphicspath{{./images}}
\usepackage[mediumspace,mediumqspace,squaren]{SIunits}
\usepackage{csquotes}

\usepackage[capitalise,nameinlink,sort&compress]{cleveref}
\crefname{equation}{}{}
\crefname{appendix}{Appendix}{appendix}
\creflabelformat{equation}{#2#1#3}
\crefmultiformat{equation}{#2#1#3}{, #2#1#3}{, #2#1#3}{, #2#1#3}
\renewcommand\eqref[1]{(\cref{#1})}

\DeclareMathOperator{\csch}{csch}

\DeclareMathOperator{\diag}{diag}

\newcommand{\matr}[1]{\mathbf{#1}}
\DeclareMathOperator{\spectrum}{Sp}

\newcommand{\E}{{\text{e}}}
\renewcommand{\vec}[1]{\mathbf{#1}}
\newcommand{\diff}[1]{\mathrm{d}{#1}}

\newcommand{\pp}[1]{\left({#1}\right)}
\newcommand{\bb}[1]{\left[ #1 \right]}
\newcommand{\eqdef}{\vcentcolon =}
\newcommand{\pdiff}[2]{\frac{\partial{#1}}{\partial{#2}}}

\newcommand{\linepdiff}[2]{{\partial{#1}}/{\partial{#2}}}

\newcommand{\fdiff}[2]{\frac{\diff{#1}}{\diff{#2}}}

\newcommand{\ub}[2]{\underbrace{#1}_{\text{#2}}}
\newcommand\restr[2]{{
 \left.\kern-\nulldelimiterspace
 #1 
 \vphantom{\big|} 
 \right|_{#2} 
 }} 

\usepackage{amssymb} 

\title{A multiscale theory for network advection-reaction-diffusion}

\author[1]{Hadrien Oliveri\thanks{To whom correspondence should be addressed: 
\href{mailto:holiveri@mpipz.mpg.de}{holiveri@mpipz.mpg.de}}}
\affil[1]{Max Planck Institute for Plant Breeding Research, Cologne 50829, Germany}

\author[2]{Emilia Cozzolino}
\affil[2]{Dipartimento di Matematica, Università di Roma Tor Vergata, Rome 00133, Italy}

\author[3]{Alain Goriely}
\affil[3]{Mathematical Institute, University of Oxford, Oxford OX2 6GG, United Kingdom}

\begin{document}

\maketitle
 \begin{abstract}
 Mathematical network models are extremely useful to capture complex propagation processes between different regions (nodes), e.g. the spread of an infectious agent between different countries, or the transport and replication of toxic proteins across different brain regions in neurodegenerative diseases. In these models, transport is modelled at the macroscale through an operator, the so-called \textit{graph Laplacian} based on the edge properties and topology, capturing the fluxes between different nodes of the network. However, this phenomenological approach fails to take into account the physical processes taking place at the microscale within the edge. A fundamental problem is then to obtain a transport operator from mechanistic principles based on the underlying transport process. Using advection-reaction-diffusion as a generic mechanism for inter-nodal exchanges, we derive a multiscale network transport model and derive the corresponding linear transport operator at the macroscale from first principles. This \textit{effective} graph Laplacian is fully determined by the transport mechanisms along the edges at the microscale. We show that this operator correctly captures the transport, and we study its scaling properties with respect to edge length.
 \end{abstract}

\section{Introduction}

Systems where mass or information move between well-defined discrete regions connected by physical pathways can often be modelled as dynamical systems on a network. In these systems, a quantity is defined at a \textit{node}, where it follows some local dynamics, and is carried through other nodes by \textit{edges} through some form of physical transport. Applications range, among others, from mobility-driven epidemiology \citep{kuhlcomputational} to protein trafficking across the brain connectome \citep{fornari2019prion,fornari2020spatially}. While modelling the dynamics at the level of nodes is relatively straightforward, a central mathematical question is to specify the transport operator governing the nodal exchanges. 
In the simplest case, by analogy with the continuum case of diffusion, one can postulate that transport is modelled by a graph Laplacian built from the network topology and weighted by its physical properties such as edges length and diameter~\citep{gautreau2007arrival,linka2020safe,raj2012network}. However, there is no first principle dictating the particular form of the transport operator, or how it is weighted apart from satisfying global physical constraints such as mass conservation, or the Fickian property that mass is transported from high to low concentrations. Here, we look at this issue as a multiscale problem. Positing an actual physical mechanism of transport along the network edges at the microscale, we determine an \textit{effective graph Laplacian} at the macroscale.

At the macroscale, we wish to obtain a generic network advection-reaction-diffusion system by combining i) transport between node through the edges; and ii) local reaction in the nodes (e.g. autocatalysis). 
The generic form of such network model is a system of ordinary differential equations (w.r.t. time $t$) for the densities at the nodes of the form \citep{kuhlcomputational}:
\begin{equation}\label{eqn:general-form}
 \fdiff{}{t}\pp{\bm{\mathcal V}
 \bm{\rho}} = \bm{\mathcal V} \matr G(\bm{\rho}) + \matr L\bm{\rho}.
\end{equation}
Here $\bm{\rho}$ is the vector of densities (with unit mass per volume), expressing the density of the considered species across all nodes; and $\bm {\mathcal V}$ the diagonal matrix of nodal volumes so that $\bm{\mathcal V}
 \bm{\rho}$ has the dimension of a mass. The function $\matr G(\bm{\rho})$ captures the local nodal reaction which may encompass various chemical reactions, as well as natural decay or clearance \citep{brennan2024role,ahern2025}. Transport is assumed to be linear and encompasses various contributions, such as passive Fickian diffusion and active transport. Typically, this contribution is modelled through the so-called \textit{graph Laplacian} $\matr L$ built from the weighted adjacency matrix of the network \citep{putra2021braiding}.

This type of system is straightforward to analyse and solve numerically, and has proven highly effective in capturing key propagation mechanisms, as it can be readily validated and parameterized against data \citep{chaggar2025personalised}. However, from a physical perspective, a key issue is to obtain the appropriate form of both the reaction term $\matr G$ and the transport operator $\matr L$. Indeed, various authors have used different ad hoc conventions \citep{abdelnour2014network,raj2015network,pandya2017predictive,pandya2019,thompson2020protein} --- some using Laplacian matrices that violate mass balance, as pointed out by \cite{putra2021braiding}. This inconsistency is rather unsettling, considering each model aspires to describe the same underlying physical processes of diffusion and transport. 

The fundamental question being addressed here is then: given well-defined physical processes specifying reaction between species and transport along the edges (the microscale), can we define a model --- defined by both a reaction term and a transport operator in \eqref{eqn:general-form} --- at the network scale (the macroscale), from first principles? 

\begin{figure}[ht!]
 \centering
 \includegraphics[width=0.8\linewidth]{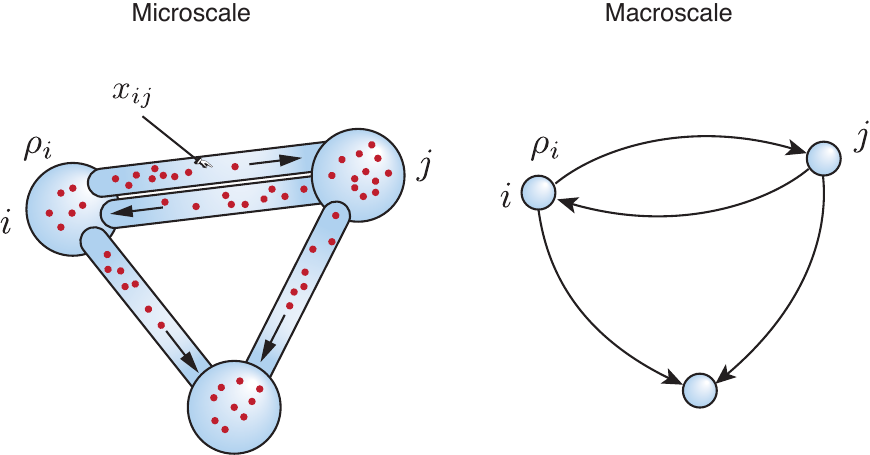}
 \caption{Schematic of the model's structure. At the microscale, the system is a physical network where nodes are compartments with volumes $\mathcal V_i$ and the edges are (one-dimensional) tubes defining a continuum domain for the advection-reaction diffusion process (with orientation of the advection shown by the arrows). At the macroscale, the system is viewed as a mathematical network of nodes and edges where the nodal densities $\rho_i$ evolve according to the macroscopic governing equation \eqref{eqn:ode-ci-final}.}
 \label{fig:fig0}
\end{figure}

To obtain the transport operator $\matr L$ and the correct form of the reaction term, we must therefore start by considering the actual physical transport within each edge at the microscale. Here, we will assume a generic advection-reaction-diffusion transport on each edge. We will show that under suitable simplifying assumptions, we can solve the microscale problem on each edge for multiple species to obtain the change of density along the edges and match the fluxes at the end of the edges to those at each node. If the density profile along an edge depends on the nodal densities in a linear fashion, one can, in principle, obtain a linear operator, which we call the \textit{effective graph Laplacian}, hence connecting the physics at the microscale to the macroscale behaviour (\cref{fig:fig0}).

\section{General setup}

A network of size $N$ is defined as a pair $\mathcal{G}=(\mathcal{V},\mathcal{E})$, where $\mathcal{V}=\{1,2,\dots,N\}$ is the set of nodes and $\mathcal{E}\subseteq \mathcal{V}\times\mathcal{V} $ is the set of edges connecting them. Since active transport between nodes is generally not symmetric, the network is directed, with edge $(i,j)$ distinct from edge $(j,i)$. An edge $(i,j)$ has arc length $\ell_{ij}$ and is assumed to have uniform cross-sectional area $w_{ij}$ so that, 
the edge volume is given by $v_{ij}\approx\ell_{ij}w_{ij}$ (without summation on repeated indices). Note that, in the case where no physical edge $(i,j)$ actually exists in the graph, we simply take $w_{ij}=0$. We first consider the advection-diffusion-reaction of $n$ species on a single edge before connecting the different edges to nodes.

\section{Advection-reaction-diffusion along an edge\label{edge}}
At the microscale, we view a single edge as a slender one-dimensional tube with arclength coordinate $s\in\bb{0,\ell}$ in which $n$ species with density $\vec x = \vec x(s,t)\in \mathbb R^n$ interact through 
\begin{equation}\label{eqn:ard-general}
 \pdiff{\vec x}{t} + \pdiff{\vec J}{s} =\matr F(\vec x). 
 \end{equation}
Here, $\matr F(\vec x)\in\mathbb R^n$ captures any source or sink, or chemical reaction between the species.
The flux $\vec J$ through the domain is given by
\begin{equation}\label{eqn:flux}
 \vec J= \matr V \vec x - \matr D \pdiff{\vec x}{s} ,
\end{equation}
with $\matr V$ the
diagonal \textit{transport velocity matrix}; and $\matr D$ the positive-definite diagonal \textit{matrix of diffusion coefficients}. Assuming that the dynamics at the nodes is much slower than that along the edges, and taking $ \matr V$ and $\matr D$ to be constant in $s$, we consider the quasi-static problem expressed by \begin{equation}\label{eqn:quasi-static-equation}
 - \matr D \vec x''+\matr V \vec x ' =\matr F(\vec x)
\end{equation}
where the apostrophe denotes differentiation w.r.t. $s$. Here, we consider Dirichlet boundary conditions of the form
\begin{equation}\label{eqn:dirichlet}
 \vec x (0) = \bm{\rho}_0(t), \quad \vec x (\ell) = \bm{\rho}_\ell(t)
\end{equation}
where the vectors $\bm{\rho}_0$ and $\bm{\rho}_\ell$ denote the (positive) densities at both ends of the domain. In a network, these will model the densities at the end nodes of a given edge, and will connect the solution of the advection-reaction-diffusion process at the edge scale to the node densities at the network scale. In general, when $\matr F$ is nonlinear, there is no closed-form solution to \eqref{eqn:quasi-static-equation} and one has to resort to numerical techniques. Here, we assume that the dominant behaviour at the microscale is governed by linear reaction and, accordingly, linearise the problem. 

Assuming the densities remain small, we linearise the reaction term $\matr F(\vec x) = \vec f_0 + \matr K \vec x + O(\lvert \vec x\rvert^2 )$ and consider the simpler problem
\begin{equation}\label{eqn:linearised-quasi-static-equation}
 - \matr D \vec x'' + \matr V \vec x' - \matr K \vec x = {\vec f_0 }.
\end{equation}
Here $\matr V$, $\matr D$ and $\matr K$ are assumed to reflect physical different processes and are independent. Hence,  
the problem \eqref{eqn:linearised-quasi-static-equation,eqn:dirichlet} is almost always non-singular and admits a unique solution. 
 Note that, for a single species ($n=1$), if $K<0$ and $f_0 \geq 0$, 
the Hopf maximum principle for elliptic equations 
implies that $x(s)\geq 0$. The same result holds in the case of multiple species if $\matr K$ is diagonal and $\diag\pp{\matr K}<0$ entry-wise. 
In the general case of a non-diagonal $\matr K$, the maximum principle for elliptic systems does not hold with the same generic conditions on the coefficients as in the case of elliptic equations \citep[see, e.g.,][]{RSMUP_1953__22__265_0}. Hence, the positivity result needs to be derived for particular classes of matrices $\matr K$. For instance, if the system is cooperative, i.e. $K_{ij}\ge 0$ for $i \ne j$ and has homogeneous Dirichlet boundary conditions, then Theorem 1.1 of \cite{DeF} guarantees positivity.

To solve \eqref{eqn:linearised-quasi-static-equation}, we observe that the system is equivalent to
\begin{equation}
 \vec z' = \matr Q \vec z - \vec q,\quad\text{with}\quad
\vec z = \bb{\begin{array}{c}
 \vec x \\
 \vec x' 
 \end{array}}, \quad \matr Q = \bb{
\begin{array}{cc}
 \vec 0 & \matr I_n \\
 - \matr D^{-1}\matr K & \matr D^{-1} \matr V 
\end{array}
 },\quad \vec q = \bb{\begin{array}{c}
 \vec 0 \\
 \matr D^{-1}\vec f_0 
 \end{array}},
\end{equation}
with $\matr I_n$ denoting the identity matrix of dimension $n\times n$.
The general solution to this problem is of the form
\begin{equation}\label{eqn:z-gen-sol}
 \vec z(s) = \E^{\matr Q s} \pp{\vec z_0 - \matr Q^{-1} \vec q} + \matr Q^{-1} \vec q,
\end{equation}
where $\vec z_0 = \vec z\pp{0}$.
Defining $\vec P = \bb{\matr I_n \,\, \matr 0}$ the $n\times 2n$ matrix which extracts the first $n$ rows of a $2n$-vector, we have
\begin{align}
 \bm{\rho}_0=\matr P \vec z_0, \quad \bm{\rho}_\ell 
 = \matr P\pp{\E^{\matr Q\ell} \vec z_0 + (\matr I_{2n} - \E^{\matr Q\ell} ) \matr Q^{-1} \vec q } .
\end{align}
These equations together form a linear system for $\vec {z_0}$:
\begin{equation}\label{eqn:z0}
 \matr M \vec z_0 = \vec k , \quad \vec k = \bb{\begin{array}{c}
  \bm{\rho}_0 \\
 \bm{\rho}_\ell + \matr P (\E^{\matr Q\ell}-\matr I_{2n} ) \matr Q^{-1} \vec q\end{array}}, \quad \matr M = \bb{\begin{array}{cc}
 \matr P \\
 \matr P \E^{\matr Q\ell}\end{array}}.
\end{equation}
If $\matr M$ is non-singular, we obtain $\vec z_0$ as a function of $\bm{\rho}_0$ and $\bm{\rho}_1$ as $\vec z_0 = \matr M^{-1} \vec k$. Replacing this result in \eqref{eqn:z-gen-sol}, the solution for $\vec z(s)$ is given by 
\begin{equation}\label{eqn:z-gen-sol2}
 \vec z(s) = \E^{\matr Q s} \pp{\matr M^{-1} \vec k - \matr Q^{-1} \vec q} + \matr Q^{-1} \vec q.
\end{equation}
From this solution we obtain the fluxes \eqref{eqn:flux}
\begin{subequations}\label{eqn:flux-0-1}
\begin{equation}
 \vec J^- \eqdef \vec J (0) = \pp{\matr V \matr P - \matr D \matr P \matr Q } \vec z_0 + \vec f_0,
\end{equation}
\begin{equation}
 \vec J^+ \eqdef \vec J (\ell) = \pp{\matr V \matr P - \matr D \matr P \matr Q } \E^{\matr Q\ell} \vec z_0 + \pp{\matr V \matr P (\matr I_{2n} - \E^{\matr Q\ell} ) + \matr D \matr P \matr Q}\E^{\matr Q\ell}\matr Q^{-1} \vec q .
\end{equation}
\end{subequations}

To summarize, from the densities at both ends of the edge, we obtain the density profile and fluxes along that edge, in the steady, linearized regime. We can now extend this solution to the whole network and describe the transport of material between nodes.

\section{Network model}

We now turn to the case of a full network made of nodes and edges which transport species between those nodes. Here the strategy is to use the solution obtained in \cref{edge} to express the fluxes between the nodes and thereby derive the effective Laplacian operator.

\subsection{Graph advection-reaction-diffusion and effective Laplacian\label{network-ard}}

We describe the dynamics of evolution of the chemical species on the graph. 
Let $x_{ij}^\mu (s)$ be the edge density of species $\mu$ on edge $\pp{i,j}$ at position $s\in\bb{0,\ell_{ij}}$. The node density at node $i$ for the species $\mu$ is noted $\rho_i^\mu$.
The advection-reaction-diffusion equations on the edge are then coupled with Dirichlet data by requiring that $x_{ij}^\mu$ is continuous across the edge-node interface
\begin{equation}
 x_{ij}^\mu(0) = \rho_i^\mu, \quad x_{ij}^\mu(\ell_{ij}) = \rho_j^\mu,
\end{equation}
and using the explicit quasi-static solution \eqref{eqn:z-gen-sol}. Thus, for any given set of node densities $\rho_i^\mu$, the density profiles along the edges are fully determined. 

As a starting point we look at the balance of mass at each node and each edge. The mass of species $\mu$ carried by edge $(i,j)\in \mathcal E$ is
\begin{equation}\label{eqn:mij-integral}
 m_{ij}^\mu = \int_0^{\ell_{ij}} w_{ij} x_{ij}^\mu (s)\, \diff s .
\end{equation} 
In our context, given the quasi-static solution \eqref{eqn:z-gen-sol2} for $x_{ij}^\mu$, we have that the mass carried by the edge is a function of the various densities in the associated nodes, 
$
 m_{ij}^\mu= m_{ij}^\mu(\rho_i^1, \dots, \rho_i^n,\rho_j^1, \dots, \rho_j^n)
$. Since the source term $F^\mu$ along the edge (noted $F_{ij}^\mu$ for this edge) does not depend explicitly on time, direct time differentiation provides
\begin{equation}\label{eqn:ode-mi}
\dot m_{ij}^\mu = \sum_{\nu=1}^n \pdiff{m_{ij}^\mu}{\rho_i^\nu} \dot \rho_i^\nu + \pdiff{m_{ij}^\mu}{\rho_j^\nu} \dot \rho_j^\nu.
\end{equation}

The mass intake at each node $i$ encompasses three types of terms: i) a bulk source ($G_i^\mu$) accounting, e.g., for local clearance, sources and other chemical reactions (local node reactions); ii) the flux terms $(J^\pm)^\mu_{ij}$ due to exchanges with other nodes (node-node fluxes); and iii) an additional flux corresponding to a net transfer from the neighbouring edge (edge-node fluxes); cf. \cite{tora2024network,BERTSCH2025113714}. In total we write
\begin{equation}\label{eqn:ode-ci}
 \fdiff{}{t}\pp{\mathcal V_i \rho_i^\mu } = \ub{\mathcal V_i G_i^\mu\pp{\rho_i^1, \dots, \rho_i^n}}{Local nodal reactions} + \ub{\pp{\sum_{j = 1 }^N w_{ji} (J^+)_{ji}^\mu - w_{ij} (J^-)_{ij}^\mu}}{Node-node fluxes} - \ub{\pp{\dot \rho_i^\mu \sum_{j=1}^N \sum_{\nu=1}^n\pp{\pdiff{m_{ij}^\nu}{\rho_i^\mu} + \pdiff{m_{ji}^\nu}{\rho_i^\mu} }}}{Edge-node fluxes}.
\end{equation}
The last term in \eqref{eqn:ode-ci} can be interpreted as a correction which accounts for the mass transfer from the node back into the edges due to the change of the Dirichlet condition. Indeed, the latter results in a change of the quasi-static solution for the edges and thus in the mass \eqref{eqn:mij-integral} carried by each edge. Thus the total mass carried by the edges varies with the boundary conditions, and this term models the corresponding change in mass at the node so that the total mass of the system is conserved (cf. \cref{mass-conservation}). The key modelling assumption is the effective separability of the variation of the mass carried by the single edge $(i,j)$ between nodes $i$ and $j$, namely the separation of the contributions $\pp{\linepdiff{m_{ij}^\nu}{\rho_i^\mu}}\dot\rho_i^\nu$ for $i$ and $\pp{\linepdiff{m_{ij}^\nu} {\rho_j^\mu}}\dot\rho_j^\nu $ for $j$. In other words, this means that the mass intake in the edge $(i,j)$ compensating for a change $\dot\rho_i^\mu$ of the boundary condition at node $i$ comes only from node $i$ itself (and not from $j$). 

In virtue of the quasi-static solutions for the edge concentrations, \eqref{eqn:z-gen-sol,eqn:flux-0-1}, the $m_{ij}^\mu$ and $(J^\pm)_{ij}^\mu$ can be written as explicit functions of the $\rho_i^\mu$. Thus, the $N\times n$ equations \eqref{eqn:ode-ci} form a closed system for the $\rho_i^\mu$. The goal now is to rewrite these equations in a canonical form revealing the structure of the effective graph Laplacian. 

From \eqref{eqn:flux-0-1}, we see that the left and right fluxes --- respectively $(J^-)_{ij}^\mu$ and $(J^+)_{ij}^\mu$ --- are affine functions of the $ \rho_i^\mu$ and $ \rho_j^\mu$, that is, they can both be written in the form
\begin{equation}\label{eqn:flux-ij}
 w_{ij} (J^\pm)_{ij}^\mu = (\omega^\pm)_{ij}^\mu + \sum_{\nu = 1 }^n (\mathbb A^\pm)_{ij}^{\mu\nu} \rho_i^\nu +(\mathbb B^\pm)_{ij}^{\mu\nu} \rho_j^\nu .
\end{equation}
Explicit expressions for the constant coefficients $(\omega^\pm)_{ij}^\mu$, $(\mathbb A^\pm)_{ij}^{\mu\nu}$ and $(\mathbb B^\pm)_{ij}^{\mu\nu}$ are straightforward to derive from \eqref{eqn:z0,eqn:flux-0-1}, however such expressions are not required to establish important properties of the Laplacian such as mass conservation, thus we omit them at this stage (explicit expressions will be considered in \cref{one-species}). The expression \eqref{eqn:flux-ij} thus allows us to rewrite the second term in the r.h.s. of \eqref{eqn:ode-mi} as
\begin{equation}\label{eqn:fluxes-laplacian}
 \sum_{j = 1 }^N w_{ji} (J^+)_{ji}^\mu - w_{ij} (J^-)_{ij}^\mu =\Omega_i^\mu+ \sum_{j=1}^N\sum_{\nu = 1 }^n\mathbb L_{ij}^{\mu\nu} \rho_j^\nu ,
\end{equation}
with the \textit{effective Laplacian tensor} $\mathbb L$ and \textit{base source} $\matr \Omega $ given by
\begin{subequations}\label{eqn:def-laplacian}
\begin{equation}\label{eqn:Ldef}
 \matr{\mathbb L}_{ij}^{\mu\nu} = (\mathbb A^+)_{ji}^{\mu\nu} - (\matr{\mathbb B}^-)_{ij}^{\mu\nu} - \delta_{ij}\sum_{k = 1 }^N \bb{(\mathbb A^-)_{ik}^{\mu\nu} - (\mathbb B^+)_{ki}^{\mu\nu}} ; 
\end{equation}
\begin{equation}\label{eqn:def-Omega}
 \Omega_i^\mu = \sum_{j = 1 }^N(\omega^+)_{ji}^\mu - (\omega^-)_{ij}^\mu.
\end{equation}
\end{subequations}
By the same linearity argument, we can rewrite \eqref{eqn:mij-integral} as
\begin{equation}\label{eqn:MN}
 m_{ij}^\mu = r^\mu_{ij}+\sum_{\nu}^n\mathbb M_{ij}^{\mu\nu} \rho_i^{\nu} + \mathbb N_{ij}^{\mu\nu} \rho_j^{\nu} .
\end{equation}
Finally, using \eqref{eqn:fluxes-laplacian,eqn:MN,eqn:def-laplacian}, assuming that the $\mathbb M_{ij}^{\mu\nu}$ and $\mathbb N_{ij}^{\mu\nu}$ are constant in time, and rearranging the terms, we rewrite \eqref{eqn:ode-ci} as 
\begin{equation}\label{eqn:ode-ci-final}
 \fdiff{}{t}\pp{\mathcal {\hat V}_i^\mu \rho_i^\mu} = \mathcal {\hat V}_i^\mu \hat{G}_i^\mu(\rho_i^1, \dots, \rho_i^n)+ \sum_{j=1}^N\sum_{\nu = 1 }^n \mathbb L_{ij}^{\mu\nu} \rho_j^\nu ,\quad \forall \pp{i,\mu}\in\bb{1,N}\times \bb{1,n},
\end{equation}
providing the canonical form of the system which extends \eqref{eqn:general-form} to the case of $n$ species. Here we have introduced the hatted quantities
\begin{subequations}
\begin{equation}\label{eqn:effectiveVi}
 \mathcal {\hat V}_i^\mu \eqdef \mathcal V_i + \sum_{j=1}^N\sum_{\nu = 1}^n \pp{\mathbb M^{\nu\mu}_{ij} + \mathbb N^{\nu\mu}_{ji}},
\end{equation}
\begin{equation}
\mathcal {\hat V}_i ^\mu\hat{G}_i^\mu\pp{\rho_i^1, \dots, \rho_i^n} \eqdef \mathcal V_i G_i^\mu\pp{\rho_i^1, \dots, \rho_i^n} + \Omega_i^\mu,
\end{equation}
\end{subequations}
which define respectively the \textit{effective volume matrix} and the \textit{effective nodal source matrix}. Note that the terms in the r.h.s. of \eqref{eqn:effectiveVi} are separate physical quantities defined independently, thus, the $ \mathcal {\hat V}_i^\mu $ are non-zero in general and the problem is non-degenerate. In the case of a single species, we can recover the standard form \eqref{eqn:general-form} by substituting the volume and source matrices $\bm {\mathcal V}$ and $\matr G$ with their effective, corrected counterpart.

Finally, we observe from \eqref{eqn:mij-integral} that if the edges are short, the load $m_{ij}^\mu$ carried by the edges should scale as $\sim w_{ij}\ell_{ij}$, i.e. like the volume of the edges. Thus, since $ \mathbb M_{ij}^{\nu\mu}=\linepdiff{m_{ij}^\nu}{\rho_i^\mu} $ and $ \mathbb N_{ij}^{\nu\mu}=\linepdiff{m_{ij}^\nu}{\rho_j^\mu}$, when the nodal volumes are large compared to the edge volumes, we can neglect the sum in \eqref{eqn:effectiveVi} and take $\mathcal {\hat V}_i^\mu = \mathcal V_i $.

\begin{figure}[ht!]
 \centering
\includegraphics[width=0.65\linewidth]{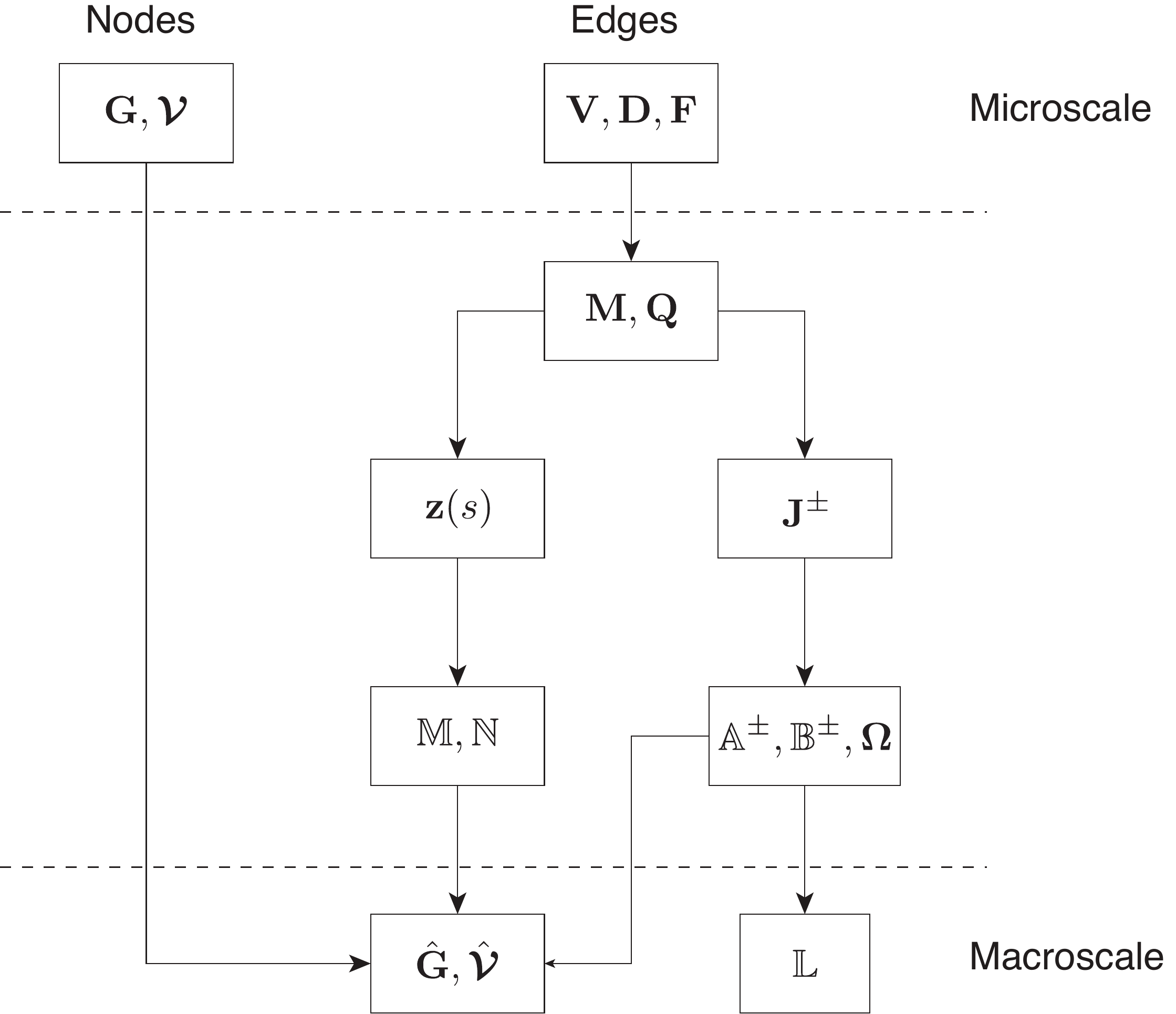}
 \caption{A summary of the upscaling process. From the initial microscale data provided by nodal and edge processes, we obtain the macroscale transport, reaction, and volume terms at the macroscale.}
 \label{algo}
\end{figure}

Since the operators introduced in this section are quite involved, we summarize the upscaling process in \cref{algo}: we started with information at the microscale concerning edge and nodal processes. Then through upscaling, we obtain at the macroscale the different operators for the evolution of the nodal concentrations. These operators are explicitly given by the properties of the system at the microscale.

\subsection{Mass conservation property of the effective Laplacian\label{mass-conservation}}

 In this section, we establish properties of the Laplacian associated with mass conservation, a key requirement for a physical transport operator (see \cref{appendix:mass-balance} for details).

 \subsubsection{General case}
The Laplacian tensor verifies the mass conservation property 
\begin{equation}\label{eqn:L-mass-balance}
 \sum_{\mu=1}^n \sum_{i=1}^N {\mathbb L_{ij}^{\mu\nu}- \mathbb A_{ji}^{\mu\nu} - \mathbb B_{ij}^{\mu\nu}} = 0.
\end{equation}
with $\mathbb A_{ij}^{\mu\nu} \eqdef(\mathbb A^+)_{ij}^{\mu\nu} - (\mathbb A^-)_{ij}^{\mu\nu}$ and $\mathbb B_{ij}^{\mu\nu} \eqdef (\mathbb B^+)_{ij}^{\mu\nu} - (\mathbb B^-)_{ij}^{\mu\nu}$ capturing the mass production along the edges (\cref{appendix:mass-balance}). This relation extends the mass conservation criterion of \cite{putra2021braiding} to multiple species and including sources along the edges. Specifically, the corrected Laplacian $\widehat{\mathbb L}_{ij}^{\mu\nu} \eqdef \mathbb L_{ij}^{\mu\nu}- \mathbb A_{ji}^{\mu\nu} - \mathbb B_{ij}^{\mu\nu}$ is mass conserving, and is obtained by removing the additional contributions of non-mass-conserving reactions along the edges.

 \subsubsection{Mass-conserving reaction} An interesting case is
\begin{equation}\label{eqn:closed}
\sum_{\mu=1}^n F_{ij}^\mu = 0 , \quad \forall (i,j)\in \mathcal E,
\end{equation}
corresponding to a scenario where species react in a way that conserves the total mass. This is the case of closed systems, including cyclic reactions (e.g. $A\rightarrow B$, $B\rightarrow C$ and $C\rightarrow A$), or contagion processes (e.g. $S+I\rightarrow I+I$). Here we show that 
\begin{equation}
 \sum_{\mu=1}^n \omega_{ij}^\mu = 0,\quad \sum_{\mu=1}^n \mathbb A_{ij}^{\mu\nu}= 0, \quad \sum_{\mu=1}^n \mathbb B_{ij}^{\mu\nu} =0,
\end{equation}
which directly yields, from \eqref{eqn:L-mass-balance},
\begin{equation}\label{eqn:cyclic-reactionb}
\sum_{i=1}^N\sum_{\mu=1}^n \mathbb L_{ij}^{\mu\nu} = 0.
\end{equation}

\subsubsection{Fully decoupled system}
In the case where the advection-reaction-diffusion system is fully decoupled, we have that $\mathbb L_{ij}^{\mu\nu}=\mathbb A_{ij}^{\mu\nu}=\mathbb B_{ij}^{\mu\nu}=0$ if $\mu\neq \nu$, thus for all $j$ and $\mu$:
\begin{equation}\label{eqn:L-mass-balance-decoupled}
 \sum_{i=1}^N {\mathbb L_{ij}^{\mu\mu}- \mathbb A_{ji}^{\mu\mu} - \mathbb B_{ij}^{\mu\mu}} = 0.
\end{equation} 
In particular for a closed system of one species with Laplacian $\matr L$ and for which $F_{ij}=0$, the mass balance property reduces to the relation $\vec 1_N \matr L = \vec 0 $ (with $\vec 1_N \eqdef \pp{1,\dots,1}\in \mathbb R^N$); cf. \cite{putra2021braiding}.

\subsection{Fick's fixed point property}

Finally, we verify that the graph Laplacian obeys the Fickian property that, in the sole presence of diffusion (i.e. in the absence of advection, conversion and sources/sinks), the uniform state $\rho_i^\mu\equiv 1$ is a fixed point. For this we require that for all $i$ and $\mu$ \eqref{eqn:ode-ci-final}
\begin{equation}\label{eqn:fick}
 \sum_{j=1}^N\sum_{\nu = 1 }^n \mathbb L_{ij}^{\mu\nu} =0. 
\end{equation}
Since the system here is decoupled, this expression simplifies to \eqref{eqn:L-mass-balance-decoupled}
\begin{equation} 
 \sum_{j=1}^N \mathbb L_{ij}^{\mu\mu} =0. 
\end{equation}
The reaction is mass-conserving, thus we also have the equality \eqref{eqn:cyclic-reactionb}
\begin{equation}
\sum_{j=1}^N \mathbb L_{ji}^{\mu\mu} = 0.
\end{equation} 
Here the Laplacian is symmetric and the two previous conditions are then equivalent \citep{putra2021braiding}, which proves the Fickian property.
\section{The case of one species\label{one-species}}

\subsection{General case}

We henceforth examine the case of a single species ($n=1$) diffusing, reacting and being advected on the network. 
In this case, the linearised advection-reaction-diffusion equation \eqref{eqn:linearised-quasi-static-equation} for one edge reduces to 
\begin{equation}
 -Dx''+Vx'-Kx = f_0.
\end{equation}
From the determinant of $\matr M$,
\begin{equation}
 \det \matr M = \frac{2 D }{\sqrt{V^2-4 DK}}\exp\pp{\frac{\ell K}{2 D}} \sinh \left(\frac{\ell \sqrt{V^2-4 DK}}{2 D}\right),
\end{equation}
we establish that this system always has a solution unless 
\begin{equation}\label{eqn:cond-sol}
\frac{\ell \sqrt{4 DK-V^2}}{2\pi D} \in \mathbb N^*.
\end{equation}
In particular, in the advection-dominated regime $V^2 > 4DK$, the square root is imaginary and solvability is always guaranteed. This loss of solvability at discrete values is not surprising and is also found in other linear systems subject to Dirichlet constraints (consider, e.g.,  
 the impossible scenario of a pendulum required to be in two different positions before and after a full-period swing). In our case, we shall see that this constraint results in a limitation on the maximum edge size. 
 
Adapting the results of \cref{network-ard} to the case of one species, the graph Laplacian is built from the eight constant matrices $A_{ij}^-$, $A_{ij}^+$, $ B_{ij}^-$, $B_{ij}^-$, $\omega_{ij}^+$, $\omega_{ij}^+$, $M_{ij}$ and $N_{ij}$ given by 
\begin{subequations}\label{eqn:QijBijCij}
\begin{equation}
 A_{ij}^- = \frac{w_{ij}}{2} \bb{V_{ij}+\sqrt{\Delta _{ij}} \coth \left(\frac{\ell_{ij}\sqrt{\Delta_{ij} } }{2 D_{ij}}\right)} ;
 \end{equation}
 \begin{equation}
 A_{ij}^+ = \frac{w_{ij} \sqrt{\Delta_{ij}}}{2} \exp\pp{\frac{\ell_{ij} V_{ij}}{2 D_{ij}}} \csch\left(\frac{\ell_{ij}\sqrt{\Delta_{ij}}}{2 D_{ij}}\right) 
 \end{equation}
 \begin{equation}
 B_{ij}^-= -\frac{w_{ij}\sqrt{\Delta_{ij} }}{2} \exp\pp{-\frac{\ell_{ij} V_{ij}}{2 D_{ij}}} \csch\left(\frac{\ell_{ij}\sqrt{\Delta_{ij} } }{2 D_{ij}}\right);
 \end{equation}
 \begin{equation}
B_{ij}^+=\frac{w_{ij}}{2} \bb{V_{ij}-\sqrt{\Delta_{ij} } \coth \left(\frac{\ell_{ij}\sqrt{\Delta_{ij} } }{2 D_{ij}}\right)};
 \end{equation}
 \begin{equation}
 \omega_{ij}^- = \frac{w_{ij}f_{0,ij} }{2 K_{ij}}\bb{\sqrt{\Delta_{ij} } \frac{\exp\pp{{\sqrt{\Delta_{ij} } \ell_{ij}}/{D_{ij}}}-2 \exp\pp{{\ell_{ij} \left(\sqrt{\Delta_{ij} }-V_{ij}\right)}/{2 D_{ij}}}+1 }{\exp\pp{{\sqrt{\Delta_{ij} } \ell_{ij}}/{D_{ij}}}-1}-V_{ij}};
 \end{equation}
\begin{align}
\omega_{ij}^+ &= -\frac{w_{ij}f_{0,ij} }{2 K_{ij}}\bb{\sqrt{\Delta_{ij} }\,\frac{\exp\pp{{\sqrt{\Delta_{ij} } \ell_{ij}}/{D_{ij}}}-2 \exp\pp{{\ell_{ij} \left(\sqrt{\Delta_{ij} }+V_{ij}\right)}/{2 D_{ij}}}+1 }{\exp\pp{{\sqrt{\Delta_{ij} } \ell_{ij}}/{D_{ij}}}-1}+V_{ij}};
\end{align}
\begin{equation}
 M_{ij} = -\frac{w_{ij} }{2 K_{ij}}\bb{\sqrt{\Delta_{ij} }\frac{\exp\pp{{\sqrt{\Delta _{ij}} \ell_{ij}}/{D_{ij}}}-2 \exp\pp{{\ell_{ij} \left(\sqrt{\Delta_{ij} }+V_{ij}\right)}/{2 D_{ij}}}+1 }{\exp\pp{{\sqrt{\Delta_{ij} } \ell_{ij}}/{D_{ij}}}-1}+V_{ij}};
\end{equation}
\begin{equation}
 N_{ij}= -\frac{w_{ij}}{2K_{ij}}\bb{\sqrt{\Delta_{ij} } \frac{\exp\pp{{\sqrt{\Delta_{ij} } \ell_{ij}}/{D_{ij}}}-2 \exp\pp{{\ell_{ij} \left(\sqrt{\Delta_{ij} }-V_{ij}\right)}/{2 D_{ij}}}+1 }{\exp\pp{{\sqrt{\Delta_{ij} } \ell_{ij}}/{D_{ij}}}-1 }-V_{ij}};
\end{equation}
\end{subequations}
with 
$\Delta_{ij} \eqdef {V_{ij}^2 - 4D_{ij}K_{ij}}$. 

To build intuition, we consider a minimal network made of two identical nodes connected via two identical but oppositely oriented edges. The Laplacian tensor for this network is a $2\times 2$ matrix that can be computed easily using \eqref{eqn:Ldef,eqn:QijBijCij}. The spectrum of this matrix characterises the scaling of the Laplacian w.r.t. to the various parameters, in particular the edge length $\ell$:
\begin{align}
\spectrum \pp{\matr L} =&\left\{-
w \sqrt{V^2 - 4 K D} \left(\cosh \left(\frac{V\ell}{2 D}\right)+\cosh \left(\frac{\ell \sqrt{V^2-4 K D}}{2 D}\right)\right) \csch\left(\frac{\ell \sqrt{V^2-4 K D}}{2 D}\right), \nonumber \right.\\ 
& \left. w \sqrt{V^2-4 K D} \left(\cosh \left(\frac{V\ell}{2 D}\right)-\cosh \left(\frac{\ell \sqrt{V^2 - 4 K D}}{2 D}\right)\right) \csch\left(\frac{\ell \sqrt{V^2 - 4 K D}}{2 D}\right)\right\}.
\end{align}
In particular, for short edges (that is for $\ell\ll D/V$), we find the approximation
\begin{equation}
\spectrum \pp{\matr L} \approx \left\{-4 Dw\ell^{-1},\, K w \ell \right\}.
\end{equation}
As can be seen, this limit case supports the choice of the so-called \textit{ballistic scaling} \citep{putra2021braiding} where the edge weights are taken proportional to $\ell^{-1}$. Conversely, it challenges the alternative quadratic scaling in $\ell^{-2}$ which has been proposed based on the analogy with finite differences \citep{thompson2020protein}. While previous studies have imposed these scalings constitutively into the Laplacian operator, our systematic approach derives the appropriate scaling directly from first principles and provides a means to assess their domain of validity. Moreover, it shows strong agreement with empirical data reported in \citep{chaggar2025personalised}.

For long edges ($\ell\gg D/V$), we have
\begin{equation}\label{eqn:long-edges}
 \spectrum \pp{\matr L} \approx \left\{- w \sqrt{V^2 - 4 KD} ,\, - w \sqrt{V^2 - 4 KD} \right\},
\end{equation}
i.e., the dependency on edge length vanishes.

\subsection{Diffusion-dominated case\label{pure-diffusion}}
It is interesting to examine the purely diffusive case with no reaction or advection. Here 
the solution established in \cref{edge} must be replaced by
\begin{equation}
  x_{ij}(s) = \frac{s}{\ell_{ij}} \pp{\rho_j - \rho_i} + \rho_i,
\end{equation}
with the fluxes at the boundaries given by
\begin{equation}
  J_{ij}^- = J_{ij}^+ = \frac{D_{ij}\pp{\rho_i - \rho_j}}{\ell_{ij}}.
\end{equation}
This solution yields the effective Laplacian \eqref{eqn:flux-ij,eqn:def-laplacian}
\begin{equation}\label{eqn:pure-diffusion}
  L_{ij} = \frac{D_{ij}w_{ji}}{\ell_{ji}} + \frac{D_{ji}w_{ij}}{\ell_{ij}} - \delta_{ij} \sum_{k=1}^N \pp{\frac{D_{ik}w_{ik}}{\ell_{ik}} + \frac{D_{ki}w_{ki}}{\ell_{ki}}}.
\end{equation}
Strikingly, the previous expression corresponds exactly to the standard form of the Laplacian with ballistic weighting used by previous authors; see \cite{putra2021braiding}. From the mass carried by an edge 
\begin{equation}
  m_{ij} = \frac{1}{2} \ell_{ij}w_{ij} \pp{\rho_j + \rho_i}, 
\end{equation}
 we further obtain
\begin{equation}\label{eqn:pure-diffusion-MN}
  M_{ij} = N_{ij} = \frac{\ell_{ij}w_{ij}}{2} .
\end{equation}
 This justifies mechanistically the usage of network models based on the Laplacian which are appropriate to capture diffusion within the network edges, provided that the Laplacian is properly weighted.

\section{Application to the Fisher-Kolmogorov-Petrovsky-Piskunov equation}

As an example, we examine the case of the Fisher-Kolmogorov-Petrovsky-Piskunov (henceforth FKPP) equation, which --- among others --- has been used as a minimal model for replication and diffusion of proteins across the brain connectome \citep{fornari2019prion}.

\subsection{General problem}

The FKPP model is introduced through the logistic reaction term at the nodes:
\begin{equation}
 G (\rho) = r^*\rho\pp{1-\epsilon^* \rho}.
\end{equation}
where $r^* $ and $\epsilon^*$ are constants which measure the growth rate and saturation in the nodes, respectively.
Along the edges, we posit \eqref{eqn:quasi-static-equation}
\begin{equation}
 - D x_{ij}'' + V x_{ij}' = r {x_{ij}} \pp{1-\epsilon x_{ij} } ,
\end{equation}
where we have assumed that $r$, $D$, $V$ and $K$ are uniform across all edges for simplicity. Choosing $\sqrt{D/r}$ and $1/r$ as reference length and time units respectively, we derive 
\begin{equation}\label{eqn:edge-fkpp}
 -  x_{ij}'' + \beta x_{ij}' = {x_{ij}} \pp{1- \epsilon x_{ij}}
\end{equation}
with 
$\beta\eqdef V/\sqrt{Dr}$ a dimensionless parameter.
We see immediately that for this system, $f_0 = 0$ and $K =1$. Since $K>0$, thus the maximum principle discussed in \cref{edge} does not apply here and the solution may be non-monotonous, and in principle negative. Using \eqref{eqn:QijBijCij}, we obtain the 
eight constant matrices
\begin{subequations}
\label{eqn:QijBijCij-FKPP}
\begin{equation}
 A_{ij}^- = \frac{w_{ij}}{2} \bb{\beta+\sqrt{\beta^2 - 4} \coth \left(\frac{\ell_{ij}\sqrt{\beta^2 - 4} }{2}\right)} ,\end{equation}
 \begin{equation}
 A_{ij}^+ = \frac{w_{ij}\sqrt{\beta^2 - 4}}{2} \exp\pp{\frac{\beta \ell_{ij}}{2}} \csch\left(\frac{\ell_{ij}\sqrt{\beta^2-4}}{2} \right),\end{equation}
 \begin{equation}
 B_{ij}^-= -\frac{w_{ij}\sqrt{\beta^2 - 4}}{2} \exp\pp{-\frac{\beta\ell_{ij} }{2 }} \csch\left(\frac{\ell_{ij}\sqrt{\beta^2 - 4} }{2 }\right),\end{equation}\begin{equation}
B_{ij}^+= \frac{w_{ij}}{2} \bb{\beta-\sqrt{\beta^2 - 4} \coth \left(\frac{\ell_{ij}\sqrt{\beta^2 - 4} }{2}\right)},\end{equation}
\begin{equation}\label{eqn:Mij-1species}
 M_{ij} = -\frac{w_{ij}}{2 }\bb{\sqrt{\beta^2-4 }\frac{\exp\pp{\ell_{ij}\sqrt{\beta^2-4}}-2 \exp\pp{{\ell_{ij} \left(\sqrt{\beta^2-4 }+\beta\right)}/2}+1}{\exp\pp{\ell_{ij}\sqrt{\beta^2-4}}-1}+\beta},\end{equation}\begin{equation}\label{eqn:Nij-1species}
 N_{ij}= -\frac{w_{ij}}{2}\bb{\sqrt{\beta^2-4} \frac{\exp\pp{\ell_{ij}\sqrt{\beta^2-4}}-2 \exp\pp{{\ell_{ij} \left(\sqrt{\beta^2-4}-\beta\right)}/2}+1}{\exp\pp{{\ell_{ij}\sqrt{\beta^2-4}}}-1}-\beta},
\end{equation}
\begin{equation}
 \omega_{ij}^- = 0,\quad
\omega_{ij}^+ = 0.
\end{equation}
\end{subequations}
The Laplacian then results explicitly from \eqref{eqn:def-laplacian}:
\begin{subequations}\label{eqn:def-laplacian2}
\begin{align}\label{eqn:Ldef2}
L_{ij} &= \frac{\sqrt{\beta^2 - 4}}{2} \bb{w_{ji} \exp\pp{\frac{\beta \ell_{ji}}{2}} \csch\left(\frac{\ell_{ji}\sqrt{\beta^2-4}}{2} \right) + w_{ij} \exp\pp{-\frac{\beta\ell_{ij} }{2 }} \csch\left(\frac{\ell_{ij}\sqrt{\beta^2 - 4} }{2 }\right)} \nonumber \\ -&
\frac{\delta_{ij}}{2} \sum_{k=1}^N \bb{w_{ik} \pp{\sqrt{\beta^2 - 4} \coth \left(\frac{\ell_{ik}\sqrt{\beta^2 - 4} }{2}\right)+\beta} + w_{ki} \pp{\sqrt{\beta^2 - 4} \coth \left(\frac{\ell_{ki}\sqrt{\beta^2 - 4} }{2}\right)-\beta}}
,
\end{align}
\begin{equation}\label{eqn:def-Omega2}
 \Omega_i = 0.
\end{equation}
\end{subequations} 

In the case where $\beta \rightarrow 0$, i.e. $V\ll \sqrt{Dr}$, \eqref{eqn:def-laplacian2,eqn:Mij-1species,eqn:Nij-1species} reduce to the remarkably simple formulae
\begin{equation}
\begin{gathered}\label{eqn:Ldef3}
L_{ij} =  w_{ij} \csc \ell_{ij} + w_{ji} \csc {\ell_{ji}} - 
\delta_{ij} \sum_{k=1}^N \pp{w_{ik}  \cot \ell_{ik} + w_{ki} \cot \ell_{ki} }, \\
 \Omega_{i} = 0,\quad M_{ij} = N_{ij} = w_{ij}\tan\pp{\ell_{ij}/2}.
\end{gathered}
\end{equation}

Note that the solvability condition \eqref{eqn:cond-sol} requires that $\ell < 2\pi / \sqrt{4 -\beta^2} $. Near this threshold the linear approximation blows up and fails to provide an accurate approximation as it misses the nonlinear saturation effect. Past this threshold, the solution is non-positive and, thus, ceases to be physical. For small $\beta < 2$, we thus must enforce $\ell \ll 2\pi / \sqrt{4 -\beta^2}$. In the case where $\beta \geq 2$ (e.g. when the growth rate $r$ is small), $\sqrt{4 -\beta^2}$ is imaginary and solvability is universally guaranteed by \eqref{eqn:cond-sol}.

Overall, we obtain the following system
\begin{equation}\label{eqn:ode-ci-final2}
  \pp{\mathcal V_i + \sum_{j=1}^N \pp{M_{ij} + N_{ji}}} \dot\rho_i = \mathcal {V}_i r^*_i \rho_i (1-\epsilon^*_i\rho_i) + \sum_{j=1}^N L_{ij} \rho_j ,\quad \forall i \in\mathcal V.
\end{equation}

\subsection{Weakly-nonlinear solution\label{weakly-nonlinear}}

The linearised FKPP model presented previously misses the quadratic saturation effect ($\epsilon$) in \eqref{eqn:edge-fkpp}. To make progress, it is useful to consider the case of a small saturation $\epsilon\ll 1$. Through regular asymptotic expansion carried out to $O(\epsilon)$, we obtain a closed-form solution which depends quadratically on the $\rho_i$ and $\rho_j$:
\begin{equation}
  x_{ij}(s) = \vec h_{ij}(s) \cdot \bm{\rho}_{ij} + \epsilon\bm{\rho} _{ij}^\top \matr H_{ij}(s) \bm{\rho}_{ij} + O(\epsilon^2),
\end{equation} 
with $\bm{\rho}_{ij}=\pp{\rho_i,\rho_j}$; and 
where $\vec h_{ij}$ and $\matr H_{ij}$ are respectively a vector and matrix function of $s$, for which we derive a rather involved explicit expression (not shown here). The flux follows from \eqref{eqn:flux}:
\begin{equation}
  w_{ij}J_{ij}(s) = \vec u_{ij}(s) \cdot \bm{\rho}_{ij} + \epsilon \bm{\rho}^\top_{ij} \matr U_{ij}(s) \bm{\rho}_{ij} ,
\end{equation} 
with 
\begin{equation}
  \quad
  \vec u_{ij}(s)\eqdef {w_{ij}\beta\vec h_{ij}(s)-w_{ij}\vec h'_{ij}(s)},\quad \matr U_{ij} (s)\eqdef {w_{ij}\beta\matr H_{ij}(s) - w_{ij}\matr H'_{ij}(s)} .
\end{equation}
In particular, we have
\begin{equation}
  w_{ij}J_{ij}^\pm = A_{ij}^\pm \rho_i + B^\pm_{ij} \rho_j + U_{ij11}^\pm \rho_i^2+U_{ij22}^\pm \rho_j^2 +\pp{U_{ij12}^\pm +U_{ij21}^\pm} \rho_i \rho_j,
\end{equation}
with $(A_{ij}^-,B_{ij}^-)=\vec u_{ij}(0)$ and $(A_{ij}^+,B_{ij}^+)=\vec u_{ij}(\ell_{ij})$; and $\matr U_{ij}^- = \matr U_{ij}(0)$ and $\matr U_{ij}^+ = \matr U_{ij}(\ell_{ij})$.
Similarly, the mass carried by an edge is given by
\begin{equation}
  m_{ij} = \int_0^{\ell_{ij}} w_{ij}x_{ij}(s)\diff s =  \bm {\mathfrak{h}}_{ij}  \cdot \bm{\rho}_{ij} + \epsilon\bm{\rho}_{ij} ^\top \matr {\mathfrak{H}_{ij}}  \bm{\rho}_{ij}, 
\end{equation}
with \begin{equation}
   \bm {\mathfrak{h}}_{ij} \eqdef {\int_0^{\ell_{ij}} w_{ij}\vec h_{ij}(s)\diff s}=\pp{M_{ij},N_{ij}},\quad \bm {\mathfrak{H}} \eqdef \int_0^{\ell_{ij}} w_{ij}\matr H_{ij}(s)\diff s.
\end{equation}
Thus, starting from \eqref{eqn:ode-ci}, and combining the previous expression for all edges $(i,j)$ we obtain a higher-order system 
\begin{align}\label{eqn:wnl} 
&\hat{\mathcal V}_i(\rho_1,\dots,\rho_N) \dot \rho_i  = {\mathcal V}_i \hat G_i \pp{\rho_i} +  \sum_{j = 1 }^N \pp{L_{ij} + \epsilon\rho_i\mathfrak{U}_{ij} }\rho_j +  \epsilon\mathfrak{V}_{ij}\rho_j^2   ,
\end{align}
with
\begin{subequations}
\begin{align}
  \hat{\mathcal V}_i \eqdef \mathcal V_i + \sum_{j=1}^N M_{ij} + N_{ji} + \epsilon\bb{\pp{\mathfrak{H}_{ji12} +\mathfrak{H}_{ji21} +2 \mathfrak{H}_{ij11} }\rho_i + \pp{\mathfrak{H}_{ij12} +\mathfrak{H}_{ij21} +2 \mathfrak{H}_{ji22} }\rho_j}; 
\end{align}
\begin{equation}
  \hat{\mathcal V}_i\hat G_i(\rho_i)\eqdef  \mathcal V_i G_i \pp{\rho_i} + \epsilon \rho_i^2 \sum_{j = 1 }^N U_{ji11}^+-U_{ij22} ^- ;
\end{equation}
\begin{equation}
\mathfrak U_{ij} = {U_{ji12}^+ +U_{ji21}^+ - U_{ij12}^- -U_{ij21}^-};
\quad
\mathfrak V_{ij}\eqdef {U_{ji22}^+ - U_{ij11}^- }.
\end{equation}
\end{subequations}
Expression \eqref{eqn:wnl} corresponds to the previous system \eqref{eqn:ode-ci-final} augmented with nonlinear terms of order $\epsilon$.

\subsection{Comparison of the different models}

In this section we compare the different models of the network FKPP system. For simplicity we focus on the case $\beta=0$. We solve the full nonlinear problem \eqref{eqn:edge-fkpp} numerically on every edge and update the node densities accordingly following a forward Euler integration scheme. For simplicity, instead of evaluating $\linepdiff{m_{ij}}{\rho_i}$ and $\linepdiff{m_{ij}}{\rho_j}$ in \eqref{eqn:ode-ci}, which is computationally heavy, we use the formulae \eqref{eqn:Ldef3} as an approximation. This numerical solution is compared with the linearised solution associated with the effective Laplacian \eqref{eqn:ode-ci-final2}, the purely diffusive case \eqref{eqn:pure-diffusion,eqn:pure-diffusion-MN}, and the weakly nonlinear solution \eqref{eqn:wnl}.

\begin{figure}[ht!]
  \centering
\includegraphics[width=0.8\linewidth]{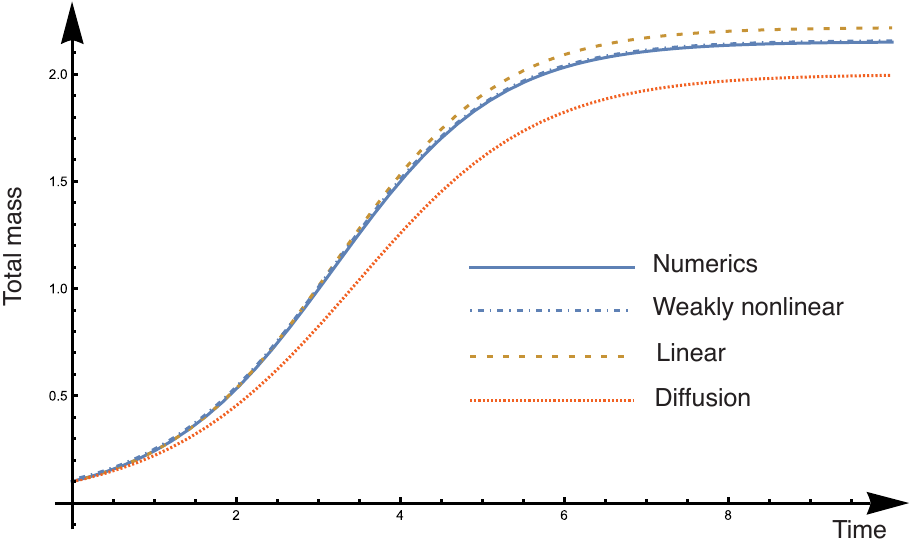}
  \caption{Example simulation of the FKPP process on a two-node network (total nodal mass). Here we compare the solution to the fully nonlinear FKPP model \eqref{eqn:edge-fkpp} computed numerically; the linearised version with effective Laplacian \eqref{eqn:Ldef3}; the weakly nonlinear problem \eqref{eqn:wnl}; and the purely diffusive problem \eqref{eqn:pure-diffusion,eqn:pure-diffusion-MN}. Parameters: $\epsilon=0.25$, $\epsilon^*=1$, $r^*=1$, $\ell=1$, $w=0.1$, $\beta=0$, $\mathcal V=1$.}
  \label{fig:simu-two-nodes}
\end{figure}

An example simulation for a two-node network is shown in \cref{fig:simu-two-nodes}. Here the network densities are initialised with $\rho_0=0.1$ and $\rho_1=0$. As can be seen, for moderate saturation $\epsilon=0.25$ and edge volumes $\ell w = 0.1$, the effective Laplacian model provides a good approximation of the dynamics. The weakly nonlinear model further improves this approximation and the two curves (for the total nodal mass) are indistinguishable. 

\subsection{Application to neurodegeneration: progression on the brain connectome}

As a proof of concept, we briefly consider the example of toxic protein propagation in the context of Alzheimer's disease. Misfolded proteins such as tau or \textalpha{}-synuclein spread prion-like along axonal bundles, where they act as seeds that catalyze further misfolding, leading to exponential growth and eventual saturation within each brain region. By coarse-graining the brain into nodes representing anatomical regions connected through the connectome, the continuum FKPP model \citep{weickenmeier2018multiphysics} reduces to a network version in which the graph Laplacian governs inter-regional transport while the nonlinear term captures local aggregation \citep{fornari2019prion}. This network FKPP approach has been shown to reproduce key spatiotemporal patterns of disease progression, including Braak staging in Alzheimer's disease, by predicting sequential regional invasion from primary seeds \citep{putra2021braiding}. Its parsimony and biological interpretability make it a valuable minimal model for investigating the interplay of transport, growth, and saturation in proteinopathies, and for generating clinically relevant predictions about disease staging and progression. Most studies using the network FKPP model use a graph Laplacian built from the weighted adjacency matrix. We demonstrate qualitatively how we can reproduce the dynamics of such systems based on our multiscale approach. 

\begin{figure}[ht!]
  \centering
\includegraphics[width=\linewidth]{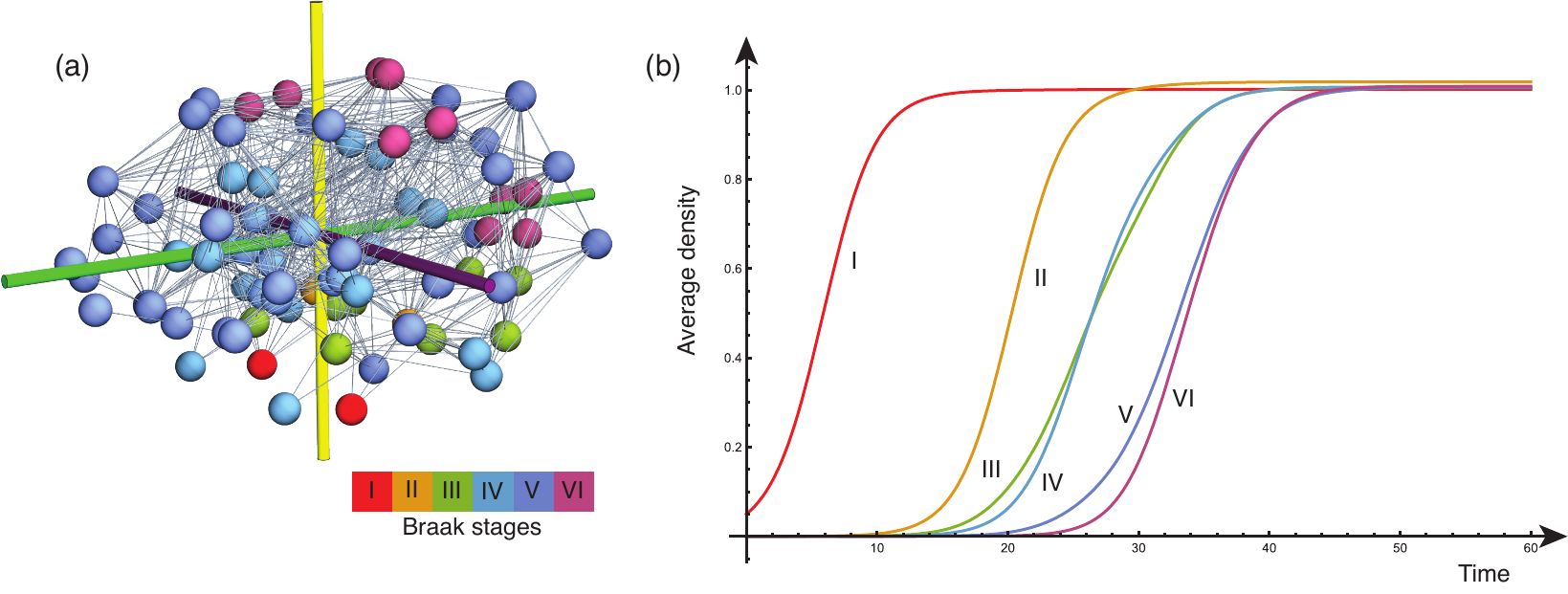}
  \caption{(a) The 83-node brain connectome and the Braak regions. (b) Example simulation of the spread of toxic proteins across the brain connectome, qualitatively in agreement with the Braak staging.}
  \label{fig:braak}
\end{figure}

The 83-node connectome shown in \cref{fig:braak}(a) is constructed by partitioning the brain into $N=83$ anatomically defined regions of interest (nodes) based on a standard brain atlas, and representing the structural connections between them as edges. The connectivity is derived from diffusion tensor magnetic resonance imaging (DTI) tractography data from 418 healthy subjects of the Human Connectome Project, aggregated into the Budapest Reference Connectome v3.0 \citep{daducci2012}. On the edges, we consider the advection-free FKPP model with $\beta=0$ \eqref{eqn:ode-ci-final2,eqn:Ldef3}. Initially, there are no toxic proteins on the network ($\rho_i=0$) except in the entorrhinal cortex (shown in red in \cref{fig:braak}), seeded with a small initial amount of proteins. This initial state is unstable and leads to a cascading propagation across the entire network \citep[parameters adapted from][]{putra2021braiding}. 

\cref{fig:braak}(b) shows the progression of the toxic load across the network, focusing on the six Braak regions, used to classify the progression of the disease, for which the average densities of toxic proteins across the region volumes are shown. As can be seen, the predicted staging of the disease is in good qualitative agreement with the Braak staging in this example. Further exploration of the parameter space can be performed following \cite{putra2021braiding}.

\section{Concluding remarks}

Using multi-compartment network models to represent spreading processes such as pathogen propagation across regions provides critical simplifications and mathematical insight, while drastically reducing computational costs, especially compared to continuum models \citep[e.g.][]{weickenmeier2018multiphysics}. These models rely on a linear operator --- typically the standard graph Laplacian --- which accounts for the exchanges between the nodes of the network. Yet, the details of this operator have been mostly phenomenological and ad hoc \citep{putra2021braiding}. A typical argument is that the diffusion scaling of the graph Laplacian (one over length squared) is usually justified as the discretisation of the diffusion operator, whereas the ballistic scaling (one over length) assumes ballistic rather than Brownian dynamics in the edges. Yet, there is little systematic or mechanistic analysis undertaken to justify these scalings properly.  Using reaction-advection-diffusion as the paradigm process by which mass moves along the edges, we derived a multiscale theory where the fluxes between the nodes result directly from the solution of a differential equation along the edges. Assuming that densities along the edges remain small, linearisation yields an explicit form for the linear transport operator --- the effective Laplacian. This operator reflects explicitly the physical quantities governing the transport process, as well as the geometry and topology of the network. Introducing the dynamics along the edges allows for the precise definition of an advection operator between nodes by accounting for the topology of the graph, thus providing a general method for constructing an active-transport process on the network \citep{Benzi10.1142}.

This method further provides a theoretical basis to study the correct Laplacian weighting. In particular it provides a justification for the length-independent weighting \eqref{eqn:long-edges} \citep{abdelnour2014network,raj2012network,raj2015network} --- which appears when the lengths of the edges are large and in the presence of active transport ($V$); and the ballistic weighting \citep{fornari2019prion,fornari2020spatially,putra2021braiding}, which captures standard diffusion (\cref{eqn:pure-diffusion}). In contrast, the quadratic weighting (one over edge's length squared) 
\citep{thompson2020protein}, which arises by canonical discretisation of the continuous Laplacian, is not captured by our method. Indeed, diffusion on a network is physically distinct from diffusion on a continuum domain, and the former may not be seen as a discretisation of the latter unless one consider a fine grid of nodes discretising the continuum, which is radically different from the basic idea of network evolution where large distances are bridged by edges.  

 The main limitation of our method is its reliance upon the assumption of linearity of the chemical reaction along the edges, precluding simple quadratic of higher order reactions along the edges, nonlinear advection or non-Fickian diffusion. However, as illustrated in the case of the FKPP equation (\cref{weakly-nonlinear}), it is possible to expand the model asymptotically to include small higher order terms capturing saturation effects. Note also that the reactions at the level of the nodes can still be arbitrarily nonlinear, as no particular hypothesis need be made regarding the functions $G_{i}^\mu$. Indeed, the goal of our model is not to include detailed nonlinear mechanisms such as nonlinear modulation of tau-protein advection along axons in Alzheimer's disease \citep{BERTSCH2025113714,tora2024network}, but rather to exhibit dominant behaviours and key scaling relationships. This approach then provides insight into the first order behaviour of the system, allowing for exhibiting the dominant behaviour of more complex nonlinear models solved only numerically. 

Our framework bridges microscale transport mechanisms with macroscale network models, thereby providing the long-missing justification for the use of graph Laplacians in a wide range of applications. Starting from first principles, our work provides a theoretical foundation for an important class of models and paves the way for their systematic application across biological, physical, and technological networks. 

 \section*{Acknowledgments}

The authors gratefully acknowledge support from the Italian PRIN project 2022W58BJ5 (\textit{PDEs and optimal control methods in mean field games, population dynamics and multi-agent models}), CUP E53D23005910006, which funded E.C.'s visit to Oxford, during which this work was conceived. E.C. further acknowledges support from the MIUR Excellence Department Project \textit{MatMod@TOV}, awarded to the Department of Mathematics, University of Rome Tor Vergata, CUP E83C23000330006. The authors also thank Michiel Bertsch for many insightful discussions.

 \section*{Data availability}

Wolfram Mathematica notebooks are available upon request. 

\section*{Author contributions} 
All authors contributed to the study conception, design, analysis, and writing.
 
\appendix

\section{Proof of the mass conservation property\label{appendix:mass-balance}}
We demonstrate that the Laplacian-like operator conserves the total mass of the system. 
We first define the total mass carried by the nodes and the edges, respectively as 
\begin{equation}
 M_{V}^\mu \eqdef \sum_{\mu=1}^n\sum_{i=1}^N \mathcal V_i\rho_i^\mu,\quad
 M_{E}^\mu \eqdef \sum_{\mu=1}^n \sum_{i=1}^N\sum_{j=1}^N m_{ij}^\mu,
\end{equation}
where, we recall, $m_{ij}^\mu$ denotes the mass of species $\mu$ carried by the edge $(i,j)$ \eqref{eqn:mij-integral}.
In total, we define the total load of the network as 
$ M = M_{V} + M_{E}$.

By summing the $n\times N$ equations \eqref{eqn:ode-ci} over the nodes and species types we write
 \begin{subequations}
\begin{equation}
 \dot M_{V} =\sum_{\mu=1}^n\sum_{i=1}^N\pp{\mathcal V_i G_i^\mu\pp{\rho_i^1, \dots, \rho_i^n} + \sum_{j = 1 }^N w_{ji} (J^+)_{ji}^\mu - w_{ij} (J^-)_{ij}^\mu - \sum_{j = 1 }^N \sum_{\nu=1}^n \pp{\pdiff{m_{ij}^\nu}{\rho_i^\mu} + \pdiff{m_{ji}^\nu}{\rho_i^\mu} }\dot \rho_i^\mu } ,\label{eqn:Mvdot}
\end{equation}
\begin{equation}
 \dot M_{E} =\sum_{\mu=1}^n\sum_{i=1}^N\sum_{j=1}^N \sum_{\nu=1}^n \pdiff{m_{ij}^\mu}{\rho_i^\nu} \dot \rho_i^\nu + \pdiff{m_{ij}^\mu}{\rho_j^\nu} \dot \rho_j^\nu=\sum_{\mu=1}^n\sum_{i=1}^N\sum_{j=1}^N \sum_{\nu=1}^n \pp{\pdiff{m_{ij}^\nu}{\rho_i^\mu} + \pdiff{m_{ji}^\nu}{\rho_j^\mu} } \dot \rho_i^\mu.\label{eqn:Medot}
\end{equation}
 \end{subequations}
 Summing the two contributions we derive
 \begin{equation}
 \dot M =\sum_{\mu=1}^n\sum_{i=1}^N\pp{\mathcal V_i G_i^\mu\pp{\rho_i^1, \dots, \rho_i^n} + \sum_{j = 1 }^N w_{ji} (J^+)_{ji}^\mu - w_{ij} (J^-)_{ij}^\mu } .\label{eqn:Mdot}
\end{equation}
We now examine the r.h.s. which captures the transport. We start by observing that the net mass \textit{produced} within an edge can be expressed using \eqref{eqn:flux-ij} as
\begin{equation}\label{eqn:flux-balance}
 w_{ij} (J^+)_{ij}^\mu - w_{ij} (J^-)_{ij}^\mu = \int_0^{\ell_{ij}} w_{ij}F_{ij}^\mu(x_{ij}^1,\dots,x_{ij}^n) \approx \omega_{ij}^\mu+ \sum_{\nu=1}^n \mathbb A_{ij}^{\mu\nu} \rho_i^\nu + \mathbb B_{ij}^{\mu\nu} \rho_j^\nu ,
\end{equation}
with $\mathbb A_{ij}^{\mu\nu} \eqdef(\mathbb A^+)_{ij}^{\mu\nu} - (\mathbb A^-)_{ij}^{\mu\nu}$, $\mathbb B_{ij}^{\mu\nu} \eqdef (\mathbb B^+)_{ij}^{\mu\nu} - (\mathbb B^-)_{ij}^{\mu\nu}$ and $\omega_{ij}^\mu \eqdef (\omega^+)_{ij}^\mu - (\omega^-)_{ij}^\mu$. Thus, summing over all edges and species, we can rearrange the terms in the last sum in \eqref{eqn:Mdot}, then use \eqref{eqn:flux-balance} to write: 
\begin{align}\label{eqn:Phidef}
\sum_{\mu=1}^n\sum_{i=1}^N\sum_{j = 1}^N {w_{ji} (J^+)_{ji}^\mu - w_{ij} (J^-)_{ij}^\mu} &= \sum_{\mu=1}^n\sum_{i=1}^N\sum_{j = 1}^N {w_{ij} (J^+)_{ij}^\mu - w_{ij} (J^-)_{ij}^\mu}
\nonumber\\& 
= \sum_{\mu=1}^n\sum_{i=1}^N\sum_{j = 1 }^N \pp{\omega_{ij}^\mu+ \sum_{\nu=1}^n \mathbb A_{ij}^{\mu\nu} \rho_i^\nu + \mathbb B_{ij}^{\mu\nu} \rho_j^\nu }. 
\end{align}
Noting, from basic manipulation rules for sums, that
\begin{align}\label{eqn:ABrho-linear-sys}
\sum_{\mu=1}^n\sum_{i=1}^N\sum_{j = 1 }^N \sum_{\nu=1}^n & \pp{\mathbb A_{ij}^{\mu\nu} \rho_i^\nu + \mathbb B_{ij}^{\mu\nu} \rho_j^\nu } = \sum_{j = 1 }^N \sum_{\nu=1}^n \pp{\sum_{\mu=1}^n\sum_{i=1}^N \pp{\mathbb A_{ji}^{\mu\nu} + \mathbb B_{ij}^{\mu\nu} }} \rho_j^\nu,
\end{align}
then using \eqref{eqn:fluxes-laplacian,eqn:ABrho-linear-sys,eqn:Phidef} and rearranging the terms, we obtain the relation 
\begin{align}\label{eqn:mass-balance-lin-sys}
 \sum_{j=1}^N\sum_{\nu = 1 }^n \pp{\sum_{\mu=1}^n \sum_{i=1}^N \pp{\mathbb L_{ij}^{\mu\nu}- \mathbb A_{ji}^{\mu\nu} - \mathbb B_{ij}^{\mu\nu}}}\rho_j^\nu = 
 \sum_{\mu=1}^n\sum_{i=1}^N\pp{- \Omega_i^\mu + \sum_{j=1}^N \omega_{ij}^\mu}. 
\end{align}
From the definition of the $\Omega_i^\mu$ \eqref{eqn:def-Omega} we see directly that the r.h.s. actually vanishes:
\begin{equation}
\sum_{\mu=1}^n\sum_{i=1}^N\pp{- \Omega_i^\mu + \sum_{j=1}^N \omega_{ij}^\mu}=0.
\end{equation} 
Finally, since the $\rho_i^\mu$ are arbitrary in \eqref{eqn:mass-balance-lin-sys}, the $n\times N$ mass conservation properties must hold:
\begin{equation}\label{eqn:L-mass-balance2}
 \sum_{\mu=1}^n \sum_{i=1}^N {\mathbb L_{ij}^{\mu\nu}- \mathbb A_{ji}^{\mu\nu} - \mathbb B_{ij}^{\mu\nu}} = 0,
\end{equation}
proving \eqref{eqn:L-mass-balance}. Note that the previous equality can also be established from \eqref{eqn:Ldef} directly.

In the particular case where there is no net creation of mass along the edges,
\begin{equation}
\sum_{\mu=1}^n F_{ij}^\mu = 0 , \quad \forall (i,j)\in \mathcal E,
\end{equation} 
\eqref{eqn:flux-balance} imply, again since $\rho_i^\nu$ and $\rho_j^\nu$ are arbitrary, that
\begin{equation}
 \sum_{\mu=1}^n \omega_{ij}^\mu = 0,\quad \sum_{\mu=1}^n \mathbb A_{ij}^{\mu\nu}= 0, \quad \sum_{\mu=1}^n \mathbb B_{ij}^{\mu\nu} =0,
\end{equation}
whence we derive \eqref{eqn:cyclic-reactionb} directly.


\begin{thebibliography}{22}
\expandafter\ifx\csname natexlab\endcsname\relax\def\natexlab#1{#1}\fi
\providecommand{\url}[1]{\texttt{#1}}
\providecommand{\href}[2]{#2}
\providecommand{\path}[1]{#1}
\providecommand{\DOIprefix}{doi:}
\providecommand{\ArXivprefix}{arXiv:}
\providecommand{\URLprefix}{URL: }
\providecommand{\Pubmedprefix}{pmid:}
\providecommand{\doi}[1]{\href{http://dx.doi.org/#1}{\path{#1}}}
\providecommand{\Pubmed}[1]{\href{pmid:#1}{\path{#1}}}
\providecommand{\bibinfo}[2]{#2}
\ifx\xfnm\relax \def\xfnm[#1]{\unskip,\space#1}\fi
\bibitem[{Abdelnour et~al.(2014)Abdelnour, Voss and Raj}]{abdelnour2014network}
\bibinfo{author}{Abdelnour, F.}, \bibinfo{author}{Voss, H.}, \bibinfo{author}{Raj, A.}, \bibinfo{year}{2014}.
\newblock \bibinfo{title}{Network diffusion accurately models the relationship between structural and functional brain connectivity networks}.
\newblock \bibinfo{journal}{Neuroimage} \bibinfo{volume}{90}, \bibinfo{pages}{335--347}.
\newblock \URLprefix \url{https://www.sciencedirect.com/science/article/abs/pii/S1053811913012597}, \DOIprefix\doi{10.1016/j.neuroimage.2013.12.039}.
\bibitem[{Ahern et~al.(2025)Ahern, Thompson, Oliveri, Lorthois and Goriely}]{ahern2025}
\bibinfo{author}{Ahern, A.}, \bibinfo{author}{Thompson, T.B.}, \bibinfo{author}{Oliveri, H.}, \bibinfo{author}{Lorthois, S.}, \bibinfo{author}{Goriely, A.}, \bibinfo{year}{2025}.
\newblock \bibinfo{title}{{Modelling the coupling between cerebrovascular pathology and amyloid beta spreading in Alzheimer's disease}}.
\newblock \bibinfo{journal}{Proceedings of the Royal Society A: Mathematical, Physical and Engineering Sciences} \bibinfo{volume}{481}, \bibinfo{pages}{20240548}.
\newblock \URLprefix \url{https://royalsocietypublishing.org/doi/10.1098/rspa.2024.0548}, \DOIprefix\doi{10.1098/rspa.2024.0548}.
\bibitem[{Benzi et~al.(2025)Benzi, Durastante and Zigliotto}]{Benzi10.1142}
\bibinfo{author}{Benzi, M.}, \bibinfo{author}{Durastante, F.}, \bibinfo{author}{Zigliotto, F.}, \bibinfo{year}{2025}.
\newblock \bibinfo{title}{Modeling advection on distance-weighted directed networks}.
\newblock \bibinfo{journal}{Mathematical Models and Methods in Applied Sciences} \bibinfo{volume}{35}, \bibinfo{pages}{1237--1265}.
\newblock \URLprefix \url{https://doi.org/10.1142/S0218202525500162}, \DOIprefix\doi{10.1142/S0218202525500162}.
\bibitem[{Bertsch et~al.(2025)Bertsch, Cozzolino and Tora}]{BERTSCH2025113714}
\bibinfo{author}{Bertsch, M.}, \bibinfo{author}{Cozzolino, E.}, \bibinfo{author}{Tora, V.}, \bibinfo{year}{2025}.
\newblock \bibinfo{title}{Well-posedness of a network transport model}.
\newblock \bibinfo{journal}{Nonlinear Analysis} \bibinfo{volume}{253}, \bibinfo{pages}{113714}.
\newblock \URLprefix \url{https://www.sciencedirect.com/science/article/pii/S0362546X24002335}, \DOIprefix\doi{10.1016/j.na.2024.113714}.
\bibitem[{Brennan et~al.(2024)Brennan, Thompson, Oliveri, Rognes and Goriely}]{brennan2024role}
\bibinfo{author}{Brennan, G.S.}, \bibinfo{author}{Thompson, T.B.}, \bibinfo{author}{Oliveri, H.}, \bibinfo{author}{Rognes, M.E.}, \bibinfo{author}{Goriely, A.}, \bibinfo{year}{2024}.
\newblock \bibinfo{title}{The role of clearance in neurodegenerative diseases}.
\newblock \bibinfo{journal}{SIAM Journal on Applied Mathematics} \bibinfo{volume}{84}, \bibinfo{pages}{S172--S198}.
\newblock \URLprefix \url{https://doi.org/10.1137/22M1487801}, \DOIprefix\doi{10.1137/22M1487801}.
\bibitem[{Chaggar et~al.(2025)Chaggar, Vogel, Binette, Thompson, Strandberg, Mattsson-Carlgren, Karlsson, Stomrud, Jbabdi, Magon et~al.}]{chaggar2025personalised}
\bibinfo{author}{Chaggar, P.}, \bibinfo{author}{Vogel, J.}, \bibinfo{author}{Binette, A.P.}, \bibinfo{author}{Thompson, T.B.}, \bibinfo{author}{Strandberg, O.}, \bibinfo{author}{Mattsson-Carlgren, N.}, \bibinfo{author}{Karlsson, L.}, \bibinfo{author}{Stomrud, E.}, \bibinfo{author}{Jbabdi, S.}, \bibinfo{author}{Magon, S.}, et~al., \bibinfo{year}{2025}.
\newblock \bibinfo{title}{Personalised regional modelling predicts tau progression in the human brain}.
\newblock \bibinfo{journal}{PLoS Biology} \bibinfo{volume}{23}, \bibinfo{pages}{e3003241}.
\newblock \URLprefix \url{https://journals.plos.org/plosbiology/article?id=10.1371/journal.pbio.3003241}, \DOIprefix\doi{10.1371/journal.pbio.3003241}.
\bibitem[{Daducci et~al.(2012)Daducci, Gerhard, Thiran et~al.}]{daducci2012}
\bibinfo{author}{Daducci, A.}, \bibinfo{author}{Gerhard, S.}, \bibinfo{author}{Thiran, J.P.}, et~al., \bibinfo{year}{2012}.
\newblock \bibinfo{title}{{T}he {C}onnectome {M}apper: {A}n {O}pen-{S}ource {P}rocessing {P}ipeline to {M}ap {C}onnectomes with {MRI}}.
\newblock \bibinfo{journal}{PLoS One} \bibinfo{volume}{7}, \bibinfo{pages}{e48121}.
\newblock \URLprefix \url{https://journals.plos.org/plosone/article?id=10.1371/journal.pone.0048121}, \DOIprefix\doi{10.1371/journal.pone.0048121}.
\bibitem[{De~Figueiredo and Mitidieri(1990)}]{DeF}
\bibinfo{author}{De~Figueiredo, D.G.}, \bibinfo{author}{Mitidieri, E.}, \bibinfo{year}{1990}.
\newblock \bibinfo{title}{Maximum principles for linear elliptic systems}.
\newblock \bibinfo{journal}{Rendiconti dell'istituto matematico di Trieste} \bibinfo{volume}{22}, \bibinfo{pages}{36--66}.
\newblock \URLprefix \url{https://link.springer.com/chapter/10.1007/978-3-319-02856-9_21}, \DOIprefix\doi{10.1007/978-3-319-02856-9_21}.
\bibitem[{Fornari et~al.(2019)Fornari, Sch{\"a}fer, Jucker, Goriely and Kuhl}]{fornari2019prion}
\bibinfo{author}{Fornari, S.}, \bibinfo{author}{Sch{\"a}fer, A.}, \bibinfo{author}{Jucker, M.}, \bibinfo{author}{Goriely, A.}, \bibinfo{author}{Kuhl, E.}, \bibinfo{year}{2019}.
\newblock \bibinfo{title}{{Prion-like spreading of Alzheimer's disease within the brain's connectome}}.
\newblock \bibinfo{journal}{Journal of the Royal Society Interface} \bibinfo{volume}{16}, \bibinfo{pages}{20190356}.
\newblock \URLprefix \url{https://royalsocietypublishing.org/doi/full/10.1098/rsif.2019.0356}, \DOIprefix\doi{10.1098/rsif.2019.0356}.
\bibitem[{Fornari et~al.(2020)Fornari, Sch{\"a}fer, Kuhl and Goriely}]{fornari2020spatially}
\bibinfo{author}{Fornari, S.}, \bibinfo{author}{Sch{\"a}fer, A.}, \bibinfo{author}{Kuhl, E.}, \bibinfo{author}{Goriely, A.}, \bibinfo{year}{2020}.
\newblock \bibinfo{title}{Spatially-extended nucleation-aggregation-fragmentation models for the dynamics of prion-like neurodegenerative protein-spreading in the brain and its connectome}.
\newblock \bibinfo{journal}{Journal of Theoretical Biology} \bibinfo{volume}{486}, \bibinfo{pages}{110102}.
\newblock \URLprefix \url{https://www.sciencedirect.com/science/article/pii/S0022519319304710}, \DOIprefix\doi{10.1016/j.jtbi.2019.110102}.
\bibitem[{Gautreau et~al.(2007)Gautreau, Barrat and Barth{\'e}lemy}]{gautreau2007arrival}
\bibinfo{author}{Gautreau, A.}, \bibinfo{author}{Barrat, A.}, \bibinfo{author}{Barth{\'e}lemy, M.}, \bibinfo{year}{2007}.
\newblock \bibinfo{title}{Arrival time statistics in global disease spread}.
\newblock \bibinfo{journal}{Journal of Statistical Mechanics: Theory and Experiment} \bibinfo{volume}{2007}, \bibinfo{pages}{L09001}.
\newblock \URLprefix \url{https://iopscience.iop.org/article/10.1088/1742-5468/2007/09/L09001}, \DOIprefix\doi{10.1088/1742-5468/2007/09/L09001}.
\bibitem[{Kuhl(2021)}]{kuhlcomputational}
\bibinfo{author}{Kuhl, E.}, \bibinfo{year}{2021}.
\newblock \bibinfo{title}{Computational Epidemiology}.
\newblock \bibinfo{publisher}{Springer}, \bibinfo{address}{Cham}.
\newblock \URLprefix \url{https://link.springer.com/book/10.1007/978-3-030-82890-5}, \DOIprefix\doi{10.1007/978-3-030-82890-5}.
\bibitem[{Linka et~al.(2020)Linka, Rahman, Goriely and Kuhl}]{linka2020safe}
\bibinfo{author}{Linka, K.}, \bibinfo{author}{Rahman, P.}, \bibinfo{author}{Goriely, A.}, \bibinfo{author}{Kuhl, E.}, \bibinfo{year}{2020}.
\newblock \bibinfo{title}{{Is it safe to lift COVID-19 travel bans? The Newfoundland story.}}
\newblock \bibinfo{journal}{Computational Mechanics} \bibinfo{volume}{66}, \bibinfo{pages}{1081--1092}.
\newblock \URLprefix \url{https://link.springer.com/article/10.1007/s00466-020-01899-x}, \DOIprefix\doi{10.1007/s00466-020-01899-x}.
\bibitem[{Pandya et~al.(2017)Pandya, Mezias and Raj}]{pandya2017predictive}
\bibinfo{author}{Pandya, S.}, \bibinfo{author}{Mezias, C.}, \bibinfo{author}{Raj, A.}, \bibinfo{year}{2017}.
\newblock \bibinfo{title}{Predictive model of spread of progressive supranuclear palsy using directional network diffusion}.
\newblock \bibinfo{journal}{Frontiers in neurology} \bibinfo{volume}{8}, \bibinfo{pages}{692}.
\newblock \URLprefix \url{https://www.frontiersin.org/journals/neurology/articles/10.3389/fneur.2017.00692}, \DOIprefix\doi{10.3389/fneur.2017.00692}.
\bibitem[{Pandya et~al.(2019)Pandya, Zeighami, Freeze, Dadar, Collins and Raj}]{pandya2019}
\bibinfo{author}{Pandya, S.}, \bibinfo{author}{Zeighami, Y.}, \bibinfo{author}{Freeze, B.}, \bibinfo{author}{Dadar, M.}, \bibinfo{author}{Collins, D.}, \bibinfo{author}{Raj, A.}, \bibinfo{year}{2019}.
\newblock \bibinfo{title}{{P}redictive model of spread of {P}arkinson's pathology using network diffusion}.
\newblock \bibinfo{journal}{NeuroImage} \bibinfo{volume}{192}, \bibinfo{pages}{178--194}.
\newblock \URLprefix \url{https://www.sciencedirect.com/science/article/abs/pii/S1053811919301685}, \DOIprefix\doi{10.1016/j.neuroimage.2019.03.001}.
\bibitem[{Pini(1953)}]{RSMUP_1953__22__265_0}
\bibinfo{author}{Pini, B.}, \bibinfo{year}{1953}.
\newblock \bibinfo{title}{Sui sistemi di equazioni lineari a derivate parziali del secondo ordine dei tipi ellittico e parabolico}.
\newblock \bibinfo{journal}{Rendiconti del Seminario Matematico della Universit\`a di Padova} \bibinfo{volume}{22}, \bibinfo{pages}{265--280}.
\newblock \URLprefix \url{https://www.numdam.org/item/RSMUP_1953__22__265_0/}.
\bibitem[{Putra et~al.(2021)Putra, Thompson, Chaggar and Goriely}]{putra2021braiding}
\bibinfo{author}{Putra, P.}, \bibinfo{author}{Thompson, T.B.}, \bibinfo{author}{Chaggar, P.}, \bibinfo{author}{Goriely, A.}, \bibinfo{year}{2021}.
\newblock \bibinfo{title}{{Braiding Braak and Braak: Staging patterns and model selection in network neurodegeneration}}.
\newblock \bibinfo{journal}{Network Neuroscience} \bibinfo{volume}{5}, \bibinfo{pages}{929--956}.
\newblock \URLprefix \url{https://direct.mit.edu/netn/article/5/4/929/107175/Braiding-Braak-and-Braak-Staging-patterns-and}, \DOIprefix\doi{10.1162/netn_a_00208}.
\bibitem[{Raj et~al.(2012)Raj, Kuceyeski and Weiner}]{raj2012network}
\bibinfo{author}{Raj, A.}, \bibinfo{author}{Kuceyeski, A.}, \bibinfo{author}{Weiner, M.}, \bibinfo{year}{2012}.
\newblock \bibinfo{title}{A network diffusion model of disease progression in dementia}.
\newblock \bibinfo{journal}{Neuron} \bibinfo{volume}{73}, \bibinfo{pages}{1204--1215}.
\newblock \URLprefix \url{https://www.cell.com/neuron/fulltext/S0896-6273(12)00135-3}, \DOIprefix\doi{10.1016/j.neuron.2011.12.040}.
\bibitem[{Raj et~al.(2015)Raj, LoCastro, Weiner et~al.}]{raj2015network}
\bibinfo{author}{Raj, A.}, \bibinfo{author}{LoCastro, E.}, \bibinfo{author}{Weiner, M.}, et~al., \bibinfo{year}{2015}.
\newblock \bibinfo{title}{Network diffusion model of progression predicts longitudinal patterns of atrophy and metabolism in {Alzheimer's} disease}.
\newblock \bibinfo{journal}{Cell reports} \bibinfo{volume}{10}, \bibinfo{pages}{359--369}.
\newblock \URLprefix \url{https://www.cell.com/cell-reports/fulltext/S2211-1247(14)01063-8}, \DOIprefix\doi{10.1016/j.celrep.2014.12.034}.
\bibitem[{Thompson et~al.(2020)Thompson, Chaggar, Kuhl, Goriely and {the Alzheimer's Disease Neuroimaging Initiative}}]{thompson2020protein}
\bibinfo{author}{Thompson, T.B.}, \bibinfo{author}{Chaggar, P.}, \bibinfo{author}{Kuhl, E.}, \bibinfo{author}{Goriely, A.}, \bibinfo{author}{{the Alzheimer's Disease Neuroimaging Initiative}}, \bibinfo{year}{2020}.
\newblock \bibinfo{title}{Protein-protein interactions in neurodegenerative diseases: A conspiracy theory}.
\newblock \bibinfo{journal}{PLoS computational biology} \bibinfo{volume}{16}, \bibinfo{pages}{e1008267}.
\newblock \URLprefix \url{https://journals.plos.org/ploscompbiol/article?id=10.1371/journal.pcbi.1008267}, \DOIprefix\doi{10.1371/journal.pcbi.1008267}.
\bibitem[{Tora et~al.(2025)Tora, Torok, Bertsch and Raj}]{tora2024network}
\bibinfo{author}{Tora, V.}, \bibinfo{author}{Torok, J.}, \bibinfo{author}{Bertsch, M.}, \bibinfo{author}{Raj, A.}, \bibinfo{year}{2025}.
\newblock \bibinfo{title}{{A network-level transport model of tau progression in the Alzheimer's brain}}.
\newblock \bibinfo{journal}{Mathematical Medicine and Biology: A Journal of the IMA} \bibinfo{volume}{00}, \bibinfo{pages}{1--27}.
\newblock \URLprefix \url{https://doi.org/10.1093/imammb/dqaf003}, \DOIprefix\doi{10.1093/imammb/dqaf003}.
\bibitem[{Weickenmeier et~al.(2018)Weickenmeier, Kuhl and Goriely}]{weickenmeier2018multiphysics}
\bibinfo{author}{Weickenmeier, J.}, \bibinfo{author}{Kuhl, E.}, \bibinfo{author}{Goriely, A.}, \bibinfo{year}{2018}.
\newblock \bibinfo{title}{The multiphysics of prion-like diseases: progression and atrophy}.
\newblock \bibinfo{journal}{Physical Review Letters} \bibinfo{volume}{121}.
\newblock \URLprefix \url{https://journals.aps.org/prl/abstract/10.1103/PhysRevLett.121.158101}, \DOIprefix\doi{10.1103/PhysRevLett.121.158101}.

\end{thebibliography}

\end{document}